\documentclass[useAMS,usenatbib,twocolumn]{mn2e}
\usepackage{psfig,subfigure}
\usepackage{graphicx}
\begin{document}
\title{Microlensing by the Halo MACHOs with Spatially Varying Mass
Function}
\author[Sohrab Rahvar]
{Sohrab Rahvar $^{1,2}$\thanks{E-mail:rahvar@sharif.edu}\\
$^1$ Department of Physics, Sharif University of Technology,\\
P.O.Box 11365--9161, Tehran, Iran\\
$^2$ Institute for Studies in Theoretical Physics and Mathematics,
P.O.Box 19395--5531, Tehran, Iran }
\date{Received 2004}
\maketitle
\begin{abstract}
The main aim of microlensing experiments is to evaluate the mean
mass of massive compact halo objects (MACHOs) and the mass
fraction of the Galactic halo made by this type of dark matter.
Statistical analysis shows that by considering a Dirac-Delta mass
function (MF) for the MACHOs, their mean mass is about that of a
white dwarf star. This result is, however, in discrepancy with
other observations such as those of non-observed expected white
dwarfs in the Galactic halo which give rise to metal abundance,
polluting the interstellar medium by their evolution. Here we use
the hypothesis of the spatially varying MF of MACHOs, proposed by
Kerins and Evans to interpret microlensing events. In this model,
massive lenses with a lower population contribute to the
microlensing events more frequently than do dominant brown dwarfs.
This effect causes the mean mass of the observed lenses to be
larger than the mean mass of all the lenses. A likelihood analysis
is performed to find the best parameters of the spatially varying
MF that are compatible with the duration distribution of Large
Magellanic Cloud microlensing candidates of the MACHO experiment.
\end{abstract}
\begin{keywords}
galaxies: halos-dark matter.
\end{keywords}
\section{Introduction}
The rotation curves of spiral galaxies (including Milky Way) show
that this type of galaxies have dark halo structure \cite{bro01}.
One of the candidates for the dark matter in the Galactic halo may
be massive compact halo objects (MACHOs). Paczy\'nski (1986)
proposed a gravitational microlensing technique as an indirect way
of detecting MACHOs. Since his proposal, many groups began
monitoring millions of stars of the Milky Way in the directions of
the spiral arms, the Galactic bulge and the Large and Small
Magellanic Clouds (LMC \& SMC) and detected hundreds of
microlensing candidates \cite{ans04,der01,sum03,afo03}. Looking in
the direction of the LMC and SMC (which are the most important for
estimating Galactic halo MAHCOs), Exp\'erience de Recherche
d'Objets Sombres (EROS)\footnote{http://eros.in2p3.fr/} and
MACHO\footnote{http://wwwmacho.mcmaster.ca/} observers found only
a dozen of microlensing candidates \cite{las00,alc00}. The
interpretation of LMC and SMC events is based on the statistical
analysis of the distribution of the duration of the events. The
result of this analysis attribute a mean mass to MACHOs and their
mass contribution in the Galactic halo. With the standard halo
model, the mean mass of lenses is evaluated to be about half
of that of the solar mass with a 20 per cent contribution in the
Galactic halo mass.\\
The results obtained by the analysis of LMC microlensing events,
however, do not agree with other observations \cite{gat01}.
Studying the kinematics of white dwarfs that have been discovered
\cite{opp01} has shown that halo white dwarfs corresponds to $1-2$
per cent of the halo mass. Recent re-analysis of the same data
\cite{spa04,tor02} shows that this fraction is an order of
magnitude smaller than the value derived in Oppenheimer et al.
(2001). On the other hand, if there were as many white dwarfs in
the halo as suggested by the microlensing experiments, they would
increase the abundance of heavy metals via the evolution of white
dwarfs and Type I Supernova explosions \cite{can97}. The other
problem is that for the mass of the MACHOs to be in the range
proposed by microlensing observations, the initial mass function
(IMF) of MACHO progenitors of the Galactic halo should be
different from those of the disc \cite{ada96,cha96}, otherwise we
should observe at the tail of mass function (MF) a large number of
luminous stars and heavy star explosions in the Galactic halos.\\
In this study we use the hypothesis of a spatially varying MF
(instead of uniform Dirac-Delta MF for the halo MACHOs) to
interpret the LMC microlensing candidates (Kerins \& Evans 1998,
hereafter KE). The physical motivation for the hypothesis of
spatially varying MFs of MACHOs comes from baryonic cluster
formation theories \cite{ash90,car94,de95}. These models predict
the spatial variation of MF in the galactic halo in such a way
that the the inner halo comprises partly visible stars, in
association with the globular cluster population, while the outer
halo comprises mostly low-mass stars and brown dwarfs.\\
We extend the work of KE by (i) using spatially varying MF model
in the power-law halo model \cite{alc96}, including the
contribution disc, spheroid and LMC for comparison with the latest
(5 yr ) LMC microlensing data (Alcock et al. 2000); (ii) using a
statistical approach applied by Green \& Jedamzik (2002) and
Rahvar (2004) to compare the distribution of the duration of the
observed events with the galactic models; and (iii) performing a
likelihood analysis to find the best parameters of the
inhomogeneous MF model.\\
The advantage of using spatially varying MF models is that the
active mean mass of lenses as the mean mass of observed lenses is
always larger than mean mass of overall lenses. This effect is
shown by a Monte-Carlo simulation, and taking it into account may
resolve the problems with interpreting microlensing data.\\
This paper is organized as follows. In Section 2 we give a brief
account of the hypothesis of spatially varying MFs and the
galactic models used in our analysis. In Section 3 we perform a
numerical simulation to generate the expected distribution of
events, taking into account the observational efficiency of the
MACHO experiment. In Section 4 we compare the theoretical
distribution of the duration of the events with the observation.
We also perform a likelihood analysis to find the best parameters
of the MF for compatibility with the observed data. The results
are discussed in Section 5.
\section{ Matter distribution in the Galactic Models}
Spiral galaxies have three components: the halo, the disc and the
bulge. We can combine these components to build various galactic
models \cite{alc96}. In this section we give a brief account on
the power-law halo and disc models which can contribute to the LMC
microlensing events. In the second part we discuss about MFs of
MACHOs and our physical motivation for considering spatially
varying MFs.
\begin{table*}
\begin{center}
\begin{tabular}{|c|c|c|c|c|c|c|c|c|c|c}
\hline
& Model :                & $S$    & $A$   & $B$  & $C$   & $D$ & $E$ & $F$ & $G$\\
\hline\hline (1)& Description & Medium & Medium& Large& Small & E6
&
Maximal& Thick & Thick \\
(2)& $\beta$              & --    &      0 & -0.2 &   0.2 &    0&    0&    0& 0\\
(3)& $q$                  & --    &     1  & 1    &    1  & 0.71
& 1& 1& 1\\
(4)& $v_a (km/s)$         & --    & 200    & 200  & 180 & 200
& 90 & 150 & 180 \\
(5) & $R_c (kpc)$         & 5    & 5  & 5   &  5  &  5   & 20 &
25 & 20\\
(6)& $R_0 (kpc)$          & 8.5  & 8.5& 8.5 & 8.5 & 8.5& 7 & 7.9
& 7.9 \\
(7)& $\Sigma_0 (M_{\odot}/pc^{2})$ & 50 & 50 & 50 & 50 & 50 & 100
& 80 & 80 \\
(8)& $R_d (kpc)$          &    3.5 & 3.5 & 3.5 & 3.5 & 3.5 & 3.5& 3 & 3 \\
(9)& $h (kpc)$          &    0.3 & 0.3 & 0.3 & 0.3 & 0.3 & 0.3
& 1 & 1 \\
(10) &$\sigma_v (km/s)$   & 31 & 31 & 31 & 31 & 31 & 31 & 49 & 49\\
\hline
\end{tabular}
\label{allmodels}
\end{center}
\caption{The parameters of the eight Galactic models: First line
is the description of the models in terms of the disk, second line
the slope of rotation curve $(\beta = 0$ flat, $\beta<0$ rising
and $\beta>0$ falling), third line the halo flattening ($q =1$
represent spherical), fourth line $(v_a)$ the normalization
velocity, fifth line $R_c$ halo core, sixth line distance of the
sun from the center of galaxy, seventh line the local column
density of the disk ($\Sigma_0 = 50 $ for canonical disk,
$\Sigma_0 = 80$ for maximal thin disk and $\Sigma_0 = 40$ for
thick disk), eighth line disk scalelength, the ninth line disk
scalehight and tenth line is the adopted one-dimensional velocity
dispersion of disk, perpendicular to our line of sight.}
\end{table*}
\subsection{Power-law halo mode}
A large set of axisymmetric galactic halo models are the "power
law " models with a matter density distribution given by
\cite{eva94} :
\begin{eqnarray}
\rho(R,z)&=&  \frac{{V_a}^2{R_c}^{\beta}}{4\pi G
q^2}\nonumber\\
&\times& \frac{{R_c}^2(1+2q^2) + R^2(1-\beta q^2) +
z^2[2-(1+\beta)/q^2]}{({R_c}^2 + R^2 + z^2/q^2)^{(\beta+4)/2}},
\label{rho}
\end{eqnarray}
where $R$ and $z$ are the coordinates in the cylindrical system,
$R_c$ is the core radius and $q$ is the flattening parameter which
is the axial ratio of the concentric equipotentials, the parameter
$\beta$ determines whether the rotational curve asymptotically
rises, falls or is flat and the parameter $V_a$ determines the
overall depth of the potential well and hence gives the typical
velocities of objects in the halo. The dispersion velocity of
particles in the halo can
be obtained by averaging the square of the velocity over the phase space.\\
Apart from the Galactic halo, there are other components of the
Milky Way such as the Galactic disc, spheroid and LMC disc that
can contribute to the LMC microlensing events. The matter
distribution of the disc is described by double exponentials
\cite{bin87} and the MF of this structure is taken according to
the {\it Hubbel Space Telescope} (HST) observations (Gould.,
Bahcall \& Flynn 1997). The second component of the Milky Way
which may also contribute to the microlensing events is the Milky
Way Spheroid. The spheroid density is given by $\rho_{spher} =
1.18\times 10^{-4}(r/R_0)^{-3.5}M_{\odot}pc^{-3}$, where $R_0$ is
the distance of the Sun from the center of Galaxy (Guidice.,
Mollerach \& Roulet 1994; Alcock et al 1996). We take the
dispersion velocity for this structure to be $\sigma_v = 120
km/s$. The matter density of LMC disc as the third structure that
can contribute to the microlensing events is also described as a
double exponential with the parameters $R_d = 1.57 kpc$, $h = 0.3
kpc$ and $\sigma = 25 km/s$, where $R_d$ is the disc scalelength,
$h$ is the disc
scalehight and $\sigma$ is the dispersion velocity \cite{gyu00}.\\
The combination of galactic substructures as galactic models
denoted by $S, A, B, C, D, E, F$ and $G$. The parameters of these
models are given in Table. 1.
\subsection{Spatially varying mass function }
The tradition in the interpretation of gravitational microlensing
data is to use the Dirac-Delta function as the simplest MF of
Galactic halo MACHOs. Colour-magnitude diagram studies of the
population of stars in the Galactic disc, bulge and other galaxies
show that MF behaves like a power law function, where the mean
mass of the stars depends on the density of interstellar medium
where the stars have been formed. Following this argument, the MF
of MACHOs in the Galactic halo may also follow a power-law
function. Fall \& Rees (1985) proposed a cooling mechanism for the
globular cluster formation and on the same basis, the hydrogen
clouds cooling mechanism can produce a cluster of brown dwarfs
(Ashman 1990). The dependence of the mass of stars on the density
of the star formation medium may causes the heavy MACHOs produced
in the dense inner
regions and the light ones at the diluted areas of the halo boarder. \\
Here in our study we use $MF(r)=\delta[M - M(r)]$ as the simplest
spatially varying MF, proposed by KE. The mass scale $M(r)$ in
this model decreases monotonically as
$M_U(\frac{M_L}{M_U})^{r/R_{halo}}$, where $r$ is the distance
from the center of the Galaxy, $M_L$ and $M_U$ are the mass scales
representing the lower and upper limits of the mass function and
$R_{halo}$ is the size of Galactic halo contains MACHOs.
Considering the cold dark matter component for the the halo, the
size of the Galactic halo may be larger than $R_{halo}$.
\section{Microlensing Events in the Spatially Varying MF}
In this section our aim is to generate microlensing events in the
spatially varying MF and compare them with the observed data. The
overall rate of microlensing events owing to the contribution of
the halo, disk, spheroid and LMC itself is given by
\begin{equation}
\frac{d\Gamma}{dt} = f\frac{d\Gamma}{dt}(halo) +
\frac{d\Gamma}{dt}(disk) + \frac{d\Gamma}{dt}(spheroid) +
\frac{d\Gamma}{dt}(LMC),
 \label{rate}
\end{equation}
where, $f$ is the fraction of the halo made by the MACHOs. The
parameter $f$ can be obtained by comparing the observed optical
depth with that of the expected value from the galactic models
\cite{alc95}. The observed optical depth is given by $\tau_{obs} =
\frac{\pi}{4E}\Sigma t_i$ , for 13 microlensing candidates of the
MACHO experiment (see Table 3) where $E=6.12 \times 10^7$
objects-years exposure time, $\tau_{obs} = 4.43\times 10^{-8}$
\cite{alc00}. The observed optical depth is sensitive to our
estimation of the duration of the events. On the other hand, the
theoretical optical depth is given by
\begin{equation}
\tau_{expected} =
\frac{\pi}{4}\int{\frac{d\Gamma}{dt}\epsilon(t)tdt},
\end{equation}
where $\epsilon(t)$ is the observational efficiency. Table 2 shows
the results of comparison between the theoretical and observed
optical depths. We use the evaluated value of $f$ in each model to
obtain the distribution of the duration of events. Fig.
\ref{te_dist} compares the normalized distributions of the
duration of events for uniform and spatially varying MFs in eight
galactic models. \\
The advantage of using a spatially varying MF is that heavy MACHOs
in the Galactic halo contribute to gravitational microlensing
events more frequently than do dominant-light MACHOs. This effect
can be shown by a Monte-Carlo simulation. Before explaining the
simulation we introduce two parameters of the passive and active
mean masses of lenses. The passive mean mass is defined as the
mean mass of the overall lenses of the Galactic halo. This mass is
independent of the gravitational microlensing observation and can
be obtained directly by averaging over the masses of MACHOs
\begin{eqnarray}
<M>_{passive} &=&
\frac{\int\phi[M,x]M(x)d^3xdM}{\int\phi[M,x]d^3xdM},\nonumber \\
&=& \frac{\int\rho(x)d^3x}{\int\frac{\rho(x)}{M(x)}d^3x},
\label{passmass}
\end{eqnarray}
the second equation is obtained by substituting the spatially varying MF model.\\
In contrast to the passive mean mass, we define the active mean
mass of lenses as the mean mass of observed microlensing
candidates. It is clear that in the case of a uniform Dirac-Delta
MF, these two masses are equal, but in the spatially varying MF,
the active mean mass of lenses is always larger than the passive one. \\
The algorithm of our simulation for evaluating the active mean
mass of lenses is (i) selecting the position of lenses according
to the position distribution function of MACHOs along our line of
sight; (ii) calculating duration of the events $(t_e)$ and
comparing them with the observational efficiency of MACHO
experiment and (iii) calculating the mean mass of selected evens.
To select the location of a lens, we imagine that we make
observations for a given interval of $T_{obs}$. The probability
that a MACHO is located at a distance $x=\frac{D_{l}}{D_{s}}$ from
the observer playing the role of a lens, thereby magnifying one of
the background stars of the LMC, is
\begin{equation}
\frac{d\Gamma}{dx} = 4\sqrt{\frac{G
D_{s}}{M(x)c^2}x(1-x)}v_t(x)\rho(x),
\end{equation}
where $M(x)$ is the mass of the MACHO and can be substituted by
the spatially varying MF model and $v_t(x)$ is the transverse
velocity of the lens with respect to our line of sight. The
duration of the events (after picking up the location of lenses)
is obtained by $t_e = \frac{2R_E(x)}{v_t(x)}$. Each time at the
the Monte-Carlo simulation loop, by comparing the duration of the
event with the observational efficiency of the MACHO experiment,
the event is selected or rejected. The mass of selected events are
used to calculate the mean mass of observed MACHOs. Table \ref{3}
shows the results of our simulation, the passive $<M_{ml}>$ and
active $<\widetilde{M}_{ml}>$ mean mass of lenses for different
galactic models. As we expected, in all the galactic models the
active mean mass is always larger than the passive one. This means
that in spite of the light abundant brown dwarfs in the Galactic
halo, lenses with the larger masses produce
most microlensing events.\\
\section{Comparison of the microlensing candidates with the galactic models}
In this section our aim is to compare the expected events from the
spatially varying MF with the microlensing candidates. The next
step is to find the best parameters for the spatially varying MF
model which are compatible with the data. Two statistical
parameters, the width of the distribution of the duration of the
events and its mean value are used in our comparison. These
parameters are defined as follows \cite{gre02,rah04}:
\begin{figure}
\centering
\includegraphics[width=100mm]{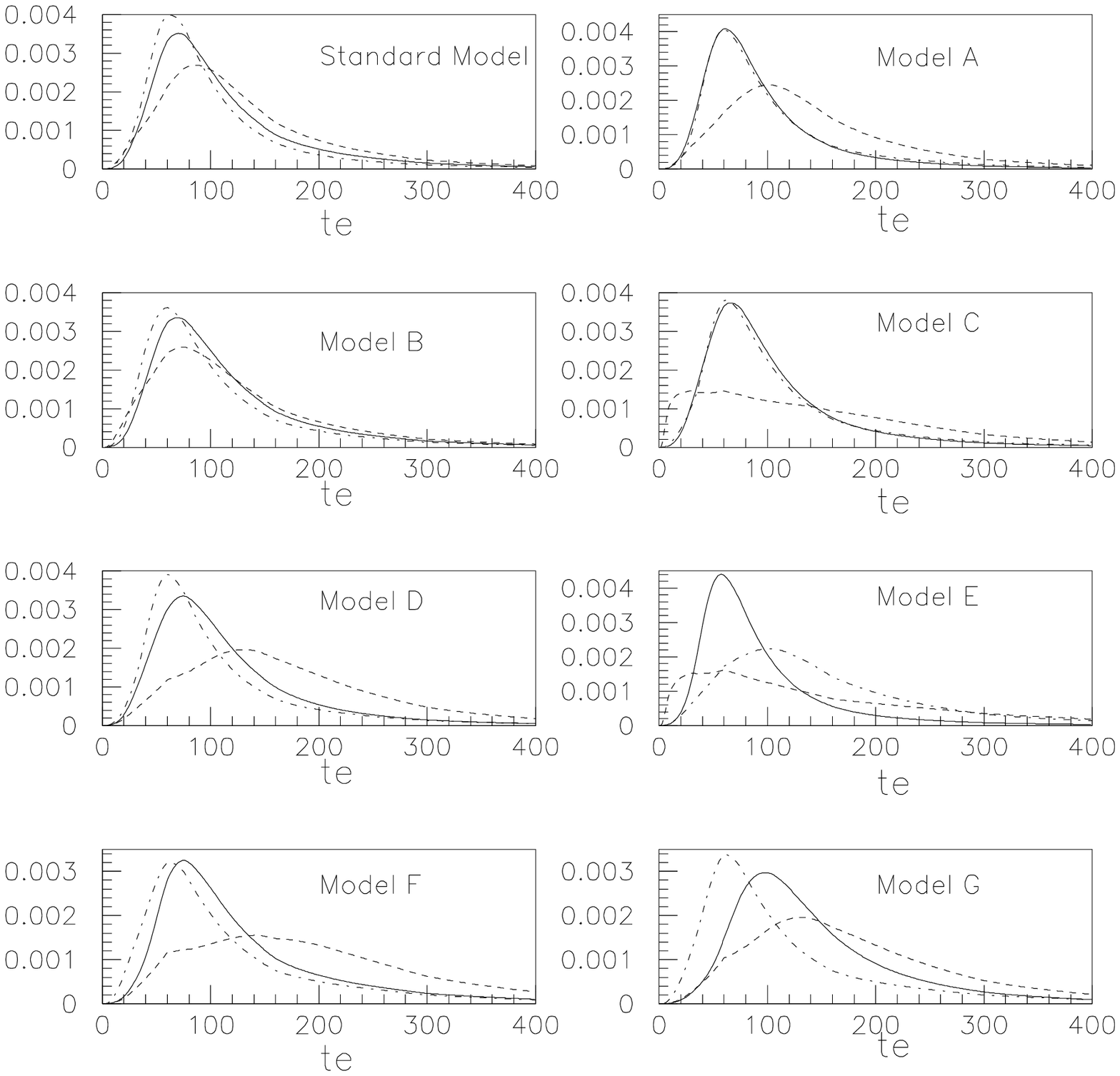}
\caption[]{The normalized distribution of events duration in the
galactic models of $Standard, A, B, C, D, E, F$ and $G$,
multiplied to the observational efficiency of MACHO experiment.
The solid and dashed lines represent the distributions of events
in the uniform and spatially varying Dirac-Delta MFs. The
dash-dotted line is resulted from the likelihood analysis for the
best parameters of KE model to be compatible with the observed
data. The duration distributions indicated by solid and
dashed-dotted are similar while the dashed-line which represents
the KE model is different than them.} \label{te_dist}
\end{figure}
\begin{figure}
\centering
\includegraphics[width=100mm]{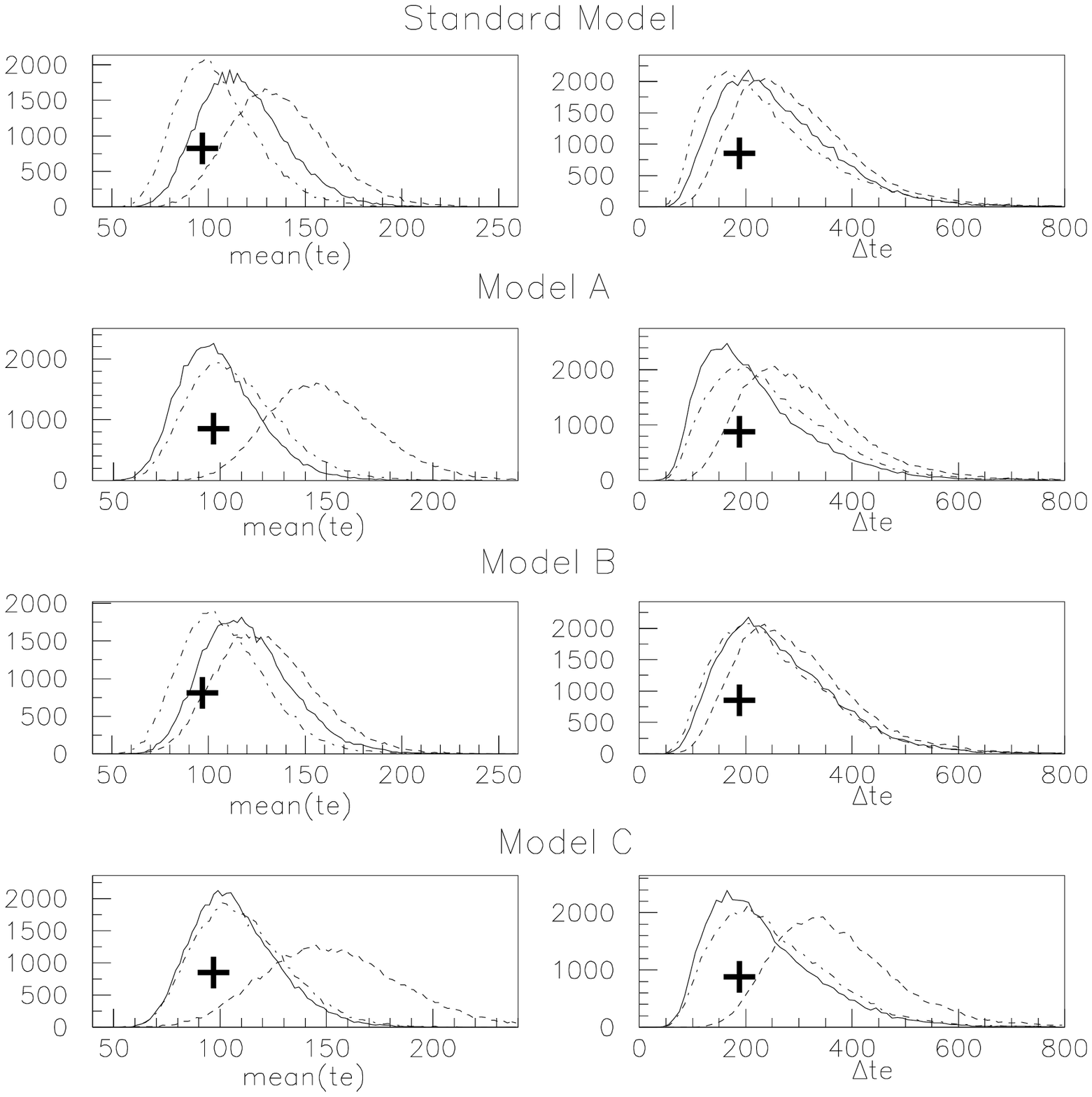}
\caption{ The expected distributions of the mean (left column) and
the width (right column) of the duration of events are shown for
the $Standard$, $A$, $B$ and $C$ Galactic models. The cross
indicates the values of $<t_e>$ and $\Delta t_e$ from the 13
microlensing candidates of MACHO experiment. These distributions
are shown for three categories of uniform Dirac Delta MF (solid
line), KE spatially varying MF (dashed line) and KE model with the
parameters derived from the likelihood analysis (dash-dotted
lines). Most of the galactic models in the spatially varying MF
with the optimized parameters are compatible with the observed
data. }\label{fig2}
\end{figure}
\begin{figure}
\centering
\includegraphics[width=100mm]{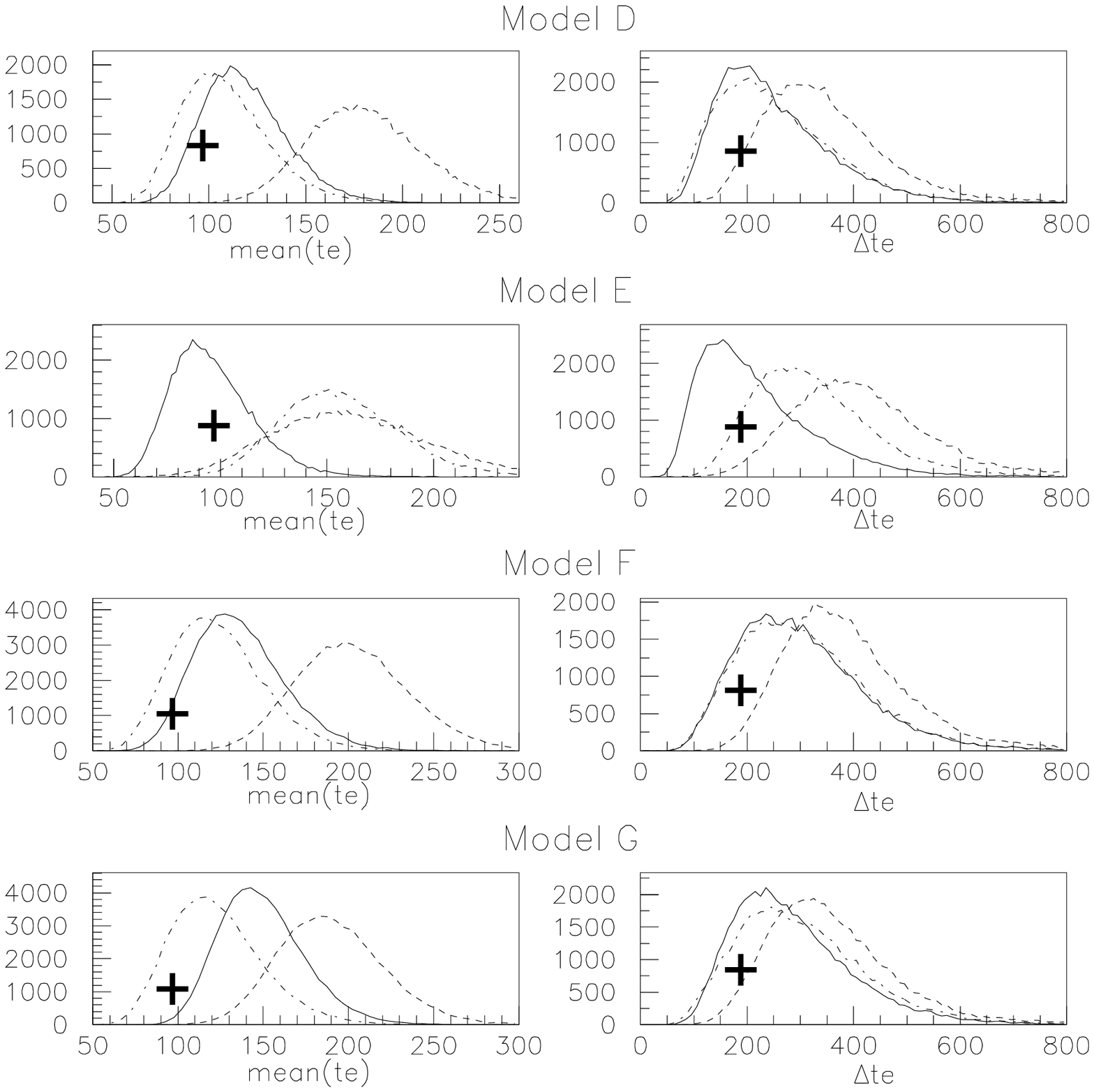}
 \caption{The expected distributions of the mean (left column) and the width
(right column) of the duration of events are shown for the  $D$,
$E$, $F$ and $G$ Galactic models. The cross indicates the values
of $<t_e>$ and $\Delta t_e$ from the 13 microlensing candidates of
MACHO experiment These distributions are shown for three
categories of uniform Dirac Delta MF (solid line), KE spatially
varying MF (dashed line) and KE model with the parameters derived
from the likelihood analysis (dash-dotted lines). Most of the
galactic model in the spatially varying MF with the optimized
parameters, except model E are compatible with the observed data.}
 \label{fig3}
\end{figure}
\begin{eqnarray}
\Delta t_e &=& Max(t_e) - Min(t_e), \\
<t_e> &=& 1/N \Sigma t_e.
\end{eqnarray}
$\Delta t_e$ and $<t_e> $ for the LMC candidates are $188$ and
$97$ d, respectively (see Table \ref{macho_data}). We perform a
Monte Carlo simulation to generate the mentioned statistical
parameters from the theoretical distribution of $t_e$ for
comparison with the observations. In this simulation we make an
ensemble of 13 microlensing events where those events are picked
up from the theoretical distribution of the duration and in each
set of events, the mean and the width of the duration of the
events are calculated. The mean of and the width from each set is
used to generate the distributions.
\begin{figure}
\centering
\includegraphics[width=100mm]{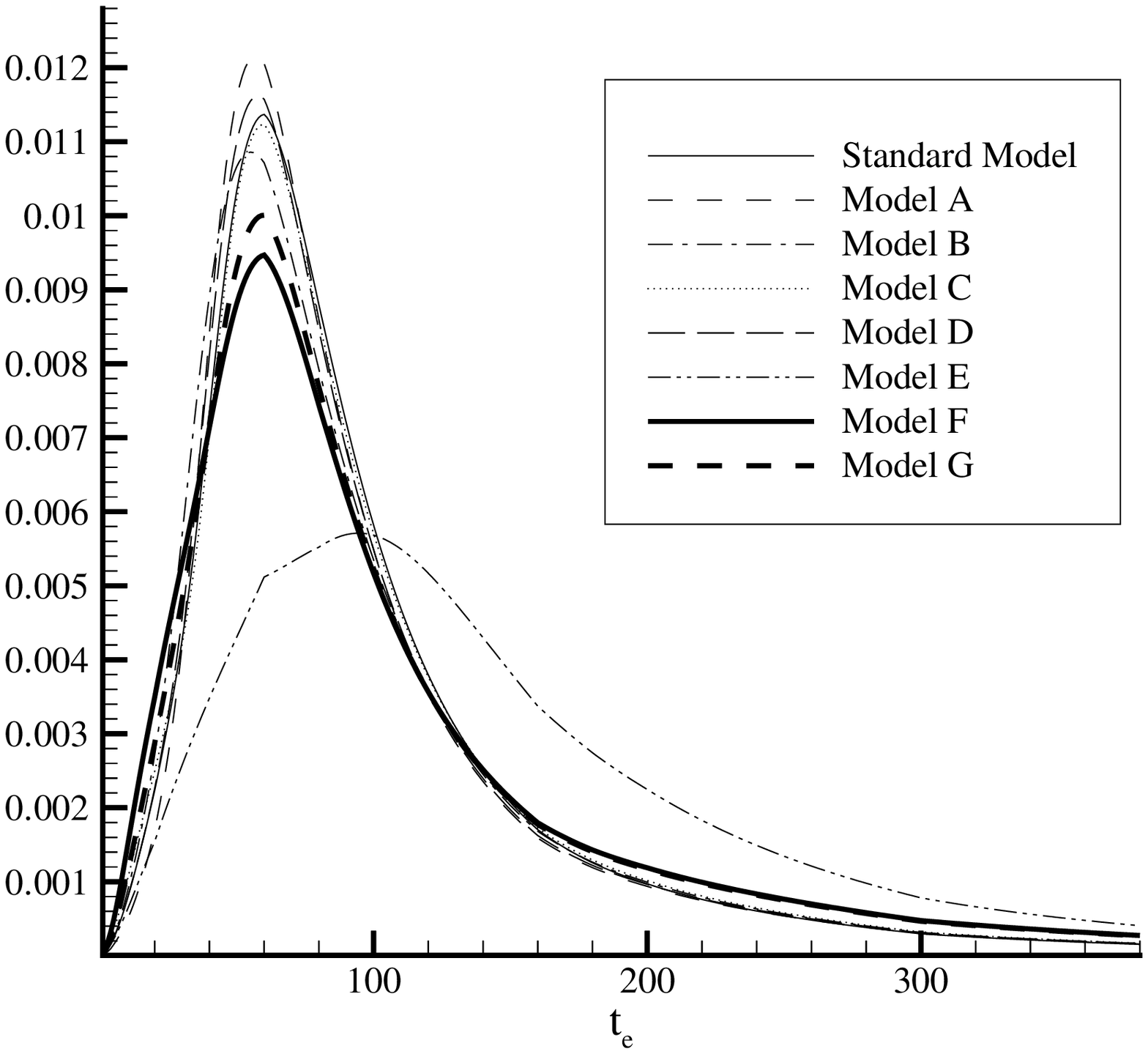}
 \caption{The duration distribution of events in eight galactic
 models with the spatially varying MF, where the MF parameters
 are optimized to be compatible with the data by the likelihood
  analysis. The expected distribution of events can be obtained by
multiplying the duration distribution to the observational
efficiency of microlensing experiment.} \label{fig4}
\end{figure}
\begin{table*}
\begin{center}
\begin{tabular}{|c|c|c|c|c|c|c|c|c|c|}
\hline
$Events (1) $  & $Model (2)$ & $MF (3)$  &$halo size (kpc) (4)$ & $M_{L}(5) $
& $M_{U} (6)$ & $<M_{ml}>  (7)$ & $<\widetilde{M}_{ml}> (8)$& $f_{ML} (9)$  \\
\hline\hline
13       & S     &  $U$           & --       & --      & --      &  0.54                     &  0.54                           &0.20   \\
13       & S     &  $KE$        &100       & $10^{-3}$    & 3       &  0.05             &  0.44               &0.16 \\
13       & S     &  $LA$          &126       & $10^{-3}$    & 1       &  0.16              &  0.26               &0.2 \\
6        & A     & $U$            & --       & --  & --     & 0.32        & 0.32              &0.41   \\
13       & A     & $KE$         &100  &  $10^{-3}$ & 3    & 0.19      & 1.05              & 0.13   \\
13       & A     & $LA$           &177  &  $10^{-3}$  & 0.5  & 0.16         & 0.31              & 0.14   \\
13       & B     & $U$            & --  & --  & --  & 0.66         & 0.66              & 0.12   \\
13       & B     & $KE$         & 100  &  $10^{-3}$  & 3    & 0.17      & 0.97                & 0.1   \\
13       & B     & $LA$          &163   & $10^{-3}$   & 0.9 &  0.22         & 0.5                & 0.1   \\
6        & C     & $U$            & --  & --  & -- & 0.21       & 0.21              & 0.61   \\
13       & C     & $KE$        & 50  &  $10^{-3}$  & 10   & 0.008        & 1.1                & 0.27  \\
13       & C     & $LA$           & 85  &  $10^{-3}$  & 0.5  & 0.04        & 0.21                & 0.25  \\
6        & D     &  $U$           & --  & --  & --   & 0.31       & 0.31              & 0.37  \\
13        & D     &  $KE$       & 100  &  $10^{-3}$  &  3  & 0.2       & 1.21                & 0.13  \\
13        & D     &  $LA$        & 103  &  $10^{-3}$  &  0.4  & 0.06         & 0.2                & 0.12  \\
6        & E     &$U$             & --  & --  & --    & 0.04       & 0.04               & 2.8  \\
13        & E     &$KE$        & 50  &  $10^{-3}$  & 10  & 0.007         & 0.31               & 1.05  \\
13        & E     &$LA$           & 87  &  $10^{-3}$  & 0.5    & 0.04       & 0.15               & 0.8  \\
13       & F     & $U$            & --  & --  & --   & 0.19       & 0.19               & 0.39  \\
13       & F     & $KE$       & 200  &  $10^{-3}$  & 2   & 0.54        & 0.99                 & 0.33  \\
13       & F     & $LA$           & 96  &  $10^{-3}$  &  0.4   & 0.04      &  0.13                 & 0.3  \\
6        & G     & $U$           & --  & --  & --  & 0.21       & 0.21                & 0.71  \\
13        & G     & $KE$     & 200  &  $10^{-3}$  & 2  & 0.56       & 1.05                & 0.18  \\
13        & G     & $LA$         & 110  &  $10^{-3}$  & 0.3  &  0.05         & 0.13                & 0.18  \\
 \hline
\end{tabular}
\end{center}
\caption{ The first column gives the number of microlensing events
that have been observed during 2 or 5.7 yrs monitoring of LMC
stars by the MACHO group. The second column indicates the name of
eight galactic models as described in Table 1. The third column
specifies the MF in each model where $U$ indicates the uniform
Dirac Delta MF which has been obtained by MACHO group, $KE$
indicates the MF proposed by Kerins and Evans (1998) and $LA$
indicates the MF which has been obtained by the likelihood
analysis. The fourth column shows the size of halo that MACHOs are
extended. The fifth column is the lower limit for the mass of
MACHOs that are located at the edge of halo and the sixth column
is the upper limit for the mass of MACHOs that reside at the
center of halo. The seventh column is the mean mass of the MACHOs
in each model, so-call passive mean mass of the lenses and the
eighth column is the active mean mass of the observed lenses by
the experiment. The ninth column shows the halo fraction made by
MACHOs in each model.} \label{3}
\end{table*}

\begin{table*}
\begin{center}
\begin{tabular}{|c|c|c|c|c|c|c|c|c|c|c|c|c|c|}
\hline
$Event :$     & 1     & 4    & 5     & 6   & 7     & 8    &13    & 14    & 15   & 18   & 21    &  23  &25
\\
\hline
$t_E$ (days)  & 34.5  & 83.3 & 109.8 & 92  & 112.6 & 66.4 &222.7 & 106.5 & 41.9 & 75.8 &141.5  & 88.9
&85.3
\\
\hline
\end{tabular}
\end{center}
\caption{Microlensing candidates observed by the MACHO experiment
during 5.7 yrs of observing 11.9 million LMC stars (Alcock et al.
2000). First row gives the name of the event according to
numbering used by the MACHO group and second row shows the
duration of event.} \label{macho_data}
\end{table*}
With this procedure we obtain the distributions of $\Delta t_e$
and $<t_e>$ for three categories of (i) Dirac-Delta MF; (ii) a
spatially varying MF; and (iii) a spatially varying MF with the
optimized parameters compatible with the data, resulting from the
likelihood analysis. Figs. \ref{fig2} and \ref{fig3} compare the
distributions of the observed $\Delta t_e$ and $<t_e>$ with three
MF models used in eight power-law galactic models. Comparing the
observed value, indicated by cross in Figs \ref{fig2} and
\ref{fig3}, with the theoretical distributions of $<t_e>$ and
$\Delta t_e$, shows that for Dirac-Delta MF, some of the galactic
models such as A, C and E are in agreement with the observations
while for the KE model non of them are compatible with
the data. \\
To find the spatially varying MF model that is compatible with the
observations, we perform a likelihood analysis to obtain the best
upper limit of the MACHO mass and the size of the halo in the KE
model. The results of analysis in the power-law galactic models
are shown in Table \ref{3} with the corresponding distribution of
the duration of events in Fig. 1. Figs. \ref{fig2} and \ref{fig3}
show that the standard model and models A, B, C and D using this
MF are in good agreement with the data.\\
In addition to the hypothesis of a spatially varying MF, there are
other hypothesis, such as self-lensing, that need to be confirmed
using sufficient statistics of microlensing events. Recent
microlensing surveys such as Optical Gravitational Lensing
Experiment (OGLE)\footnote{ http://bulge.princeton.edu/~ogle/} and
SuperMACHO\footnote{ http://www.ctio.noao.edu/~supermacho/} are
monitoring LMC stars and will provide more microlensing candidates
over the coming years. To use the results of our analysis in the
mentioned experiments, we obtained the theoretical distribution of
events in each model without applying any observational efficiency
(Fig. \ref{fig4}). The expected distribution of events in each
experiment can be obtained by
multiplying the observational efficiencies to these theoretical models.\\
It is worth to mention that our statistical analysis is sensitive
to our estimation of the duration of the microlensing candidates.
The correction with the blending effect can alter our result. The
blending effect makes a source star to be brighter than its actual
brightness, and the lensing duration appears shorter. The duration
of a microlensing event can be determined from a light curve fit
in which the brightness of the source star is included as a
fitting parameter. The main problem with this standard method is
the degeneracy caused by the fitting. High-resolution images by
the {\it HST} have been used to resolve blending by random field
stars in eight LMC microlensing events \cite{alc01}. The MACHO
group also used another procedure where each event is fitted with
a light curve that assumes no blending and then a correction is
applied to the time-scale to account for the fact that blending
tends to make the time scales appear shorter. This correction was
determined from the efficiency of a Monte Carlo simulation, and it
is a function of the time-scale of the measured event. The
procedure is designed to give the correct average event
time-scale, but it does not preserve the width of the time-scale
distribution \cite{ben04}. Green \& Jedamzik (2002) and Rahvar
(2004) showed that the width of duration of events derived from
this method is narrower than the theoretical expectations.
\section{Conclusion}
In this work we extended the hypothesis of a spatially varying MF
proposed by KE as a possible solution resolving discrepancies
between microlensing results and other observations. The main
point of this is to investigate the contradiction where
microlensing experiments predict large numbers of white dwarfs
which have not been observed.\\
The advantage of using a spatially varying MF is that we can
modify our interpretation of microlensing data. We showed that in
this model, in contrast to the Dirac-Delta MF, massive MACHOs
contribute in the microlensing events more frequently than the
low-mass ones do. To quantify our argument we defined two mass
scales, the active mean mass of MACHOs as the mean mass of lenses
that can be observed by the gravitational microlensing experiment
and the passive mean mass of MACHOs as the overall mean mass of
them. We showed that the active mean mass of MACHOs is always
larger than the passive mean mass, except in the case of a uniform
Dirac Delta MF where they are equal. \\
To test the compatibility of this model with the observed
microlensing events, we compared the duration distribution of the
events in this model with the LMC candidates of MACHO experiment.
We used two statistical parameters - the mean and the width of the
duration of events - to compare the observed data with the
theoretical models. We showed that amongst power-low halo models
some of them with a Dirac-Delta MF are compatible with the data,
while in the case KE model, almost none of them are compatible
with the data. The best parameters for the KE model were obtained
with likelihood analysis. In the spatially varying MF using the
new parameters, some Galactic models (such as standard model and
models A, B, C and D) were compatible with the data. The
hypothesis of a spatially varying MF of MACHOs may be tested by
measuring the proper motions of white dwarfs in the Galactic halo
\cite{tor02}.\\

ACKNOWLEDGMENT\\
The author thanks David Bennett for his useful comments on
blending correction of the duration of events and Sepehr Arbabi
and Mohammad Nouri-Zonoz for reading the manuscript and giving
useful comments.

\end{document}